\documentclass[11pt]{article}
\usepackage{times}
\usepackage{sidecap}
\usepackage{geometry}
\usepackage{floatrow}
\usepackage{float}
\usepackage{hyperref}
\usepackage{graphics,graphicx}
\usepackage{subcaption}
\usepackage{epsf,wrapfig}

\usepackage{natbib,bm}

\usepackage{paralist}
\let\olditem\item
\renewenvironment{thebibliography}[1]{%
  \section*{\refname}
  \let\par\relax
  \renewcommand{\item}[1][]{\olditem[\textbullet]}%
  \inparaenum}{\endinparaenum}

\newcounter{aff}

\DeclareGraphicsExtensions{.jpg,.pdf,.png,.eps}

\usepackage{caption}
\usepackage{xcolor}

\usepackage{tcolorbox}

\bibpunct{(}{)}{;}{a}{}{,}

\usepackage{soul}

\definecolor{darkcyan}{rgb}{0.0, 0.0, 0.7}

\definecolor{boxcolor}{RGB}{193, 229, 255}
\definecolor{figboxcolor}{RGB}{193, 229, 255}
\definecolor{boxcolorgray}{RGB}{227, 227, 227}
\definecolor{darkcyan}{rgb}{0.0, 0.0, 0.7}

\sethlcolor{boxcolorgray}

\def\gtrsim{\lower 2pt \hbox{$\, \buildrel {\scriptstyle >}\over
{\scriptstyle \sim}\,$}}
\def\lesssim{\lower 2pt \hbox{$\, \buildrel {\scriptstyle <}\over
{\scriptstyle \sim}\,$}}

\def\approxlt{\lower.2em\hbox{$\buildrel < \over \sim$}}
\def\approxgt{\lower.2em\hbox{$\buildrel > \over \sim$}}

\def\arcsec{$^{\prime\prime}$}

\def\kms{{\rm km\ s^{-1}}}
\def\be{\begin{equation}}
\def\ee{\end{equation}}
\def\bea{\begin{eqnarray*}}
\def\eea{\end{eqnarray*}}

\def\arcsec{^{\prime\prime}}

\def\cmc{\rm cm$^{-3}$}

\def\jref@jnl#1{{\rm#1}}
\def\aj{\jref@jnl{AJ}}			
\def\araa{\jref@jnl{ARA\&A}}		
\def\apj{\jref@jnl{ApJ}}			
\def\apjl{\jref@jnl{ApJ}}		
\def\apjs{\jref@jnl{ApJS}}		
\def\ao{\jref@jnl{Appl.~Opt.}}		
\def\apss{\jref@jnl{Ap\&SS}}		
\def\aap{\jref@jnl{A\&A}}		
\def\aapr{\jref@jnl{A\&A~Rev.}}		
\def\aaps{\jref@jnl{A\&AS}}		
\def\azh{\jref@jnl{AZh}}			
\def\baas{\jref@jnl{BAAS}}		
\def\jrasc{\jref@jnl{JRASC}}		
\def\memras{\jref@jnl{MmRAS}}		
\def\mnras{\jref@jnl{MNRAS}}		
\def\pra{\jref@jnl{Phys.~Rev.~A}}	
\def\prb{\jref@jnl{Phys.~Rev.~B}}	
\def\prc{\jref@jnl{Phys.~Rev.~C}}	
\def\prd{\jref@jnl{Phys.~Rev.~D}}	
\def\pre{\jref@jnl{Phys.~Rev.~E}}	
\def\prl{\jref@jnl{Phys.~Rev.~Lett.}}	
\def\pasp{\jref@jnl{PASP}}		
\def\pasj{\jref@jnl{PASJ}}		
\def\qjras{\jref@jnl{QJRAS}}		
\def\skytel{\jref@jnl{S\&T}}		
\def\solphys{\jref@jnl{Sol.~Phys.}}	
\def\sovast{\jref@jnl{Soviet~Ast.}}	
\def\ssr{\jref@jnl{Space~Sci.~Rev.}}	
\def\zap{\jref@jnl{ZAp}}			
\def\nat{\jref@jnl{Nature}}		
\def\iaucirc{\jref@jnl{IAU~Circ.}}
\def\aplett{\jref@jnl{Astrophys.~Lett.}}
\def\apspr{\jref@jnl{Astrophys.~Space~Phys.~Res.}}
\def\bain{\jref@jnl{Bull.~Astron.~Inst.~Netherlands}}
\def\fcp{\jref@jnl{Fund.~Cosmic~Phys.}}
\def\gca{\jref@jnl{Geochim.~Cosmochim.~Acta}}
\def\grl{\jref@jnl{Geophys.~Res.~Lett.}}
\def\jcp{\jref@jnl{J.~Chem.~Phys.}}	
\def\jgr{\jref@jnl{J.~Geophys.~Res.}}	
\def\jqsrt{\jref@jnl{J.~Quant.~Spec.~Radiat.~Transf.}}
\def\memsai{\jref@jnl{Mem.~Soc.~Astron.~Italiana}}
\def\nphysa{\jref@jnl{Nucl.~Phys.~A}}
\def\physrep{\jref@jnl{Phys.~Rep.}}
\def\physscr{\jref@jnl{Phys.~Scr}}
\def\planss{\jref@jnl{Planet.~Space~Sci.}}
\def\procspie{\jref@jnl{Proc.~SPIE}}

\def\earth{\hbox{$\oplus$}}

\def\arcsec{\hbox{$^{\prime\prime}$}}

\def\ion#1#2{#1$\;${\small\rm\@Roman{#2}}\relax}

\newcount\lecurrentfam
\def\LaTeX{\lecurrentfam=\the\fam \leavevmode L\raise.42ex
\hbox{$\fam\lecurrentfam\scriptstyle\kern-.3em A$}\kern-.15em\TeX}
\def\sizrpt{
(\fontname\the\font): em=\the\fontdimen6\font, ex=\the\fontdimen5\font
\typeout{
(\fontname\the\font): em=\the\fontdimen6\font, ex=\the\fontdimen5\font
}}
\definecolor{dark}{rgb}{ 0.34, 0.09, 0.49}

\usepackage{fancyhdr,url}
\usepackage{transparent}
\usepackage{tablefootnote}
\usepackage{mathpazo}

\begin{document}

% Title and subtitle
\noindent
\begin{raggedright}
{\Huge\bfseries Unveiling star formation at sub-solar metallicity with ALMA2040\\[0.5em]
%\Large Subtitle (if needed)
}\end{raggedright}
%\vfill
% Author block
%\thispagestyle{empty}
\noindent
%\vspace{1em}
\noindent
Francesco Fontani$^1$,
Fr\'ed\'erique Motte$^2$,
Maria T. Beltr\'an$^1$, Jan Forbrich$^3$, Jes J\o rgensen$^4$, Timea Csengeri$^5$, Laura Colzi$^6$, Víctor M. Rivilla$^6$, Giovanni Sabatini$^1$, \'Alvaro S\'anchez-Monge$^7$, Peter Schilke$^8$, Fengwei Xu$^9$,

\vspace{1em}
{
\footnotesize{
\noindent $^1$ INAF-Osservatorio Astrofisico di Arcetri, Largo E. Fermi 5, 50125 Florence, Italy \\
$^2$ Univ. Grenoble Alpes, CNRS, IPAG, 38000 Grenoble, France \\
$^3$ University of Hertfordshire, Centre for Astrophysics Research, College Lane, Hatfield AL10 9AB, UK \\
$^4$ Niels Bohr Institute, University of Copenhagen, Jagtvej 155A, 2200 Copenhagen N, Denmark. \\
$^5$ Laboratoire d’astrophysique de Bordeaux, Univ. Bordeaux, CNRS, 33615 Pessac, France \\
$^6$ Centro de Astrobiolog\'ia (CAB), CSIC-INTA, Ctra. de Ajalvir Km. 4, Torrej\'on de Ardoz, 28850 Madrid, Spain \\
$^7$ Institut de Ci\`encies de l’Espai (ICE-CSIC), Carrer de can Magrans s/n, 08193 Bellaterra, Barcelona, Spain \\
$^8$ Institute of Astrophysics, University of Cologne, Z\"ulpicher Str. 77, 50937 K\"oln, Germany, \\
$^9$ Max Planck Institute for Astronomy, K\"onigstuhl 17, 69117 Heidelberg, Germany \\
}
}
\begin{tcolorbox}[boxsep=3pt,left=0pt,right=0pt,top=1pt,bottom=1pt,colframe=boxcolor,colback=boxcolor]
\justify
\textbf{Abstract:} 
A fundamental question in astrophysics is whether the processes governing star formation are universal or depend on the physical conditions of their local galactic environment. 
The Milky Way and the Magellanic clouds offer ideal laboratories to explore this question, as the gas conditions in them span up to orders of magnitude in metallicity, gas surface density, ambient pressure, dust abundance, radiation field, and cosmic-ray flux.
In particular, both the outer Milky Way and the Magellanic clouds provide the closest low-metallicity environments with conditions similar to those of metal-poor galaxies, while remaining sufficiently close to soon allow spatial resolution of individual star-forming cores and disks. 
Recent observations with ALMA and other facilities have begun to unveil the physical and chemical properties of star-forming regions in low-metallicity environments.
Yet, their distances from the Sun still require transformational gains in current ALMA capabilities to address fundamental open questions: (1) Are the properties of prestellar cores universal? (2) Does cloud fragmentation, and the resulting stellar IMF, depend critically on metallicity? (3) How does dust evolution depend on metallicity, and what is the impact in the survival of protoplanetary disks? (4) Is the reduced amount of metals and dust shielding at low metallicity critical for the formation and survival of molecules?
%(i) How do prestellar cores form and what are their physical properties? (ii) How is the hierarchical mass distribution from clouds to cores in low-metallicity environments, and how is it linked to the Initial Mass Function? (iii) How do protoplanetary disks form and evolve, and what are their global properties? (iv) How does molecular complexity develop, and what does it reveal about the chemical pathways to stars and planets?
%With an order of magnitude improvement in sensitivity and spectral coverage, as well as enhanced {\bf imaging performances such as dynamic range, capability to image large areas and image fidelity}, 
{\bf ALMA2040 will identify similarities and differences between solar- and sub-solar-metallicity star-forming regions, and answer the question: how do different environmental conditions influence the physical and chemical pathways leading to star formation?}

\end{tcolorbox}

\newpage{}
% Text

\vspace{-0.7cm}
\section{Scientific context}
\label{context}
\vspace{-0.2cm}
Sub-solar metallicity\footnote{The typical definition of metallicity, $Z$, is ${\rm Z \equiv [Fe/H] = Log_{10}(Fe/H)_*-Log_{10}(Fe/H)_{\odot}}$, where ${\rm (Fe/H)_*}$ is the elemental abundance ratio in a cloud or star.} gas clouds or stars have an amount of elements heavier than hydrogen and helium lower than solar (i.e. $Z<Z_{\odot}$). 
In star formation, this lower abundance of metals is expected to change how gas cools, fragments, and builds stars, as heavy elements and dust grains act as the primary coolants in interstellar gas clouds.
Therefore, sub-solar metallicity clouds should have higher gas temperatures.
If core masses are determined by Jeans fragmentation, bound cores on the verge of gravitational collapse would require higher densities and/or pressure confinement in these low-metallicity, and thus warmer, clouds, thereby reducing the efficiency of forming low-mass ($M\simeq 1M_{\odot}$) cores and, ultimately, low-mass stars \citep[e.g.,][]{pes20,gjergo26}. 
This effect should be particularly strong in clusters hosting young massive stars, which have been observed to affect the brown dwarf regime of the Initial Mass Function (IMF) in solar-metallicity environments \citep{doBritodoVale2026,rom2026}.
On the side of planet formation, protoplanetary disks around sub-solar metallicity stars should contain fewer heavy solid materials, and hence be associated with a reduced efficiency in the formation of large solid planetary cores \citep{teske24,nea25}.
In addition, the reduced dust content could accelerate disk dispersal and shorten lifetimes compared to solar-metallicity regions \citep{yasui26}.
Therefore, studying regions of different metallicity is fundamental to understand the impact of environment on the process of star formation.

\begin{figure}[H]
    \centering
    \includegraphics[width=0.95\linewidth]{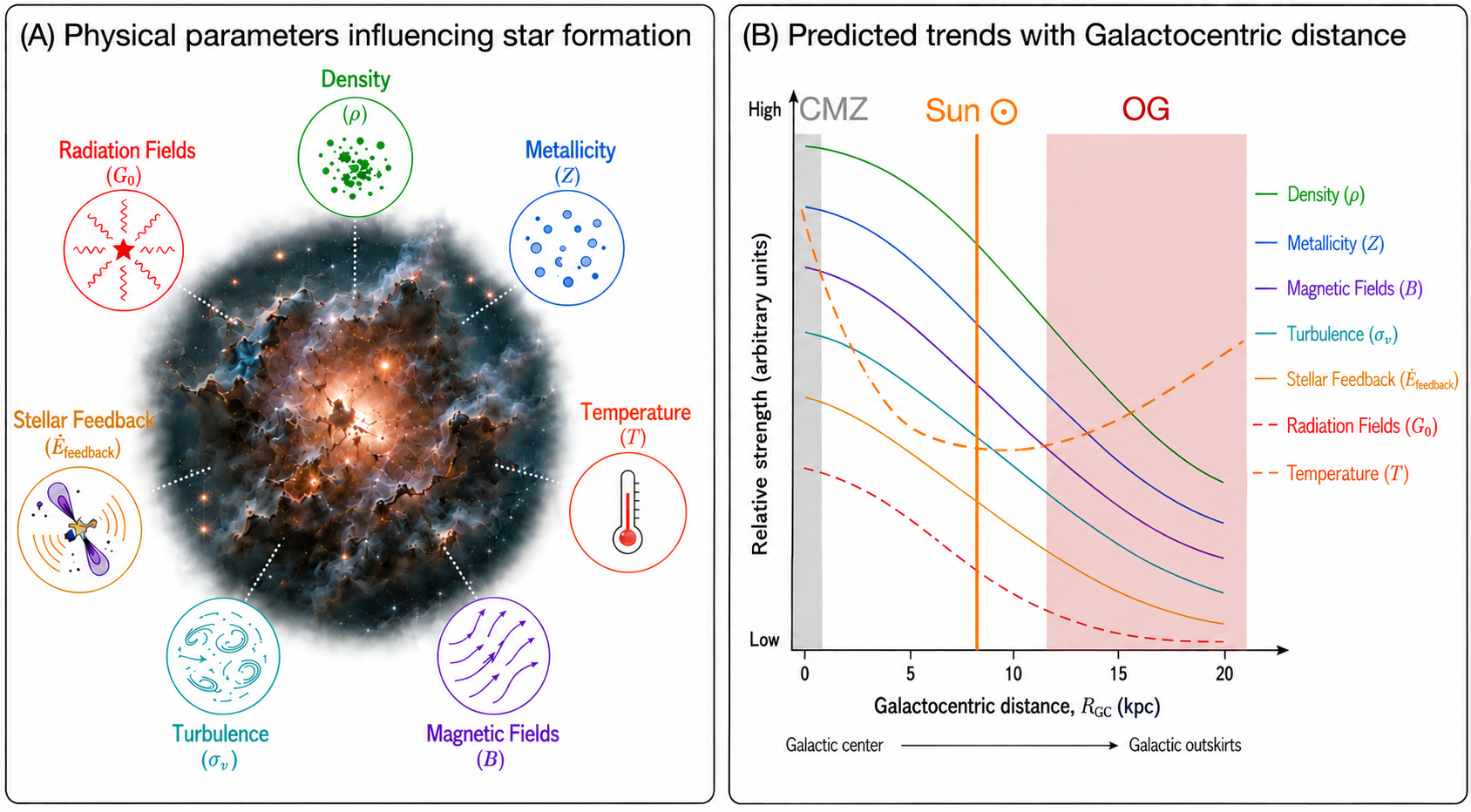}
    \caption{Key physical parameters driving star formation (A), and schematic comparison of the variations of these parameters 
    from the Central Molecular Zone (high $Z$) to the Outer (low $Z$) Galaxy (B).
    The qualitative trends shown should be interpreted only as indicative of the direction of variation, and do not imply quantitatively similar radial dependences.
    Example reference works are: Wolfire et al.~(\citeyear{wolfire03}, pressure), Ruiz-Granados et al.~(\citeyear{ruiz10}, magnetic fields), Djordjevic et al.~(\citeyear{djordjevic19}, stellar feedback), Luck \& Lambert et al.~(\citeyear{lel11}, metallicity), Bialy~(\citeyear{bialy20}, radiation fields), Braine et al.~(\citeyear{braine23}, surface density), Gigli et al.~(\citeyear{gigli25}, turbulence). 
    The Galactocentric radius of the Sun, as well as the coarse Galactocentric ranges of the Central Molecular Zone (CMZ) and the Outer Galaxy (OG), are indicated.
    The $T$ enhancement in the CMZ is due to a combination of processes, including mechanical and stellar feedback \citep[][]{henshaw23}.}
    \label{fig:comparison}
\end{figure}

The Milky Way
%, the galaxy in which the Sun was born along with hundreds of billions of similar stars, 
is a perfect laboratory to test these predicted differences, as it is 
a barred spiral with a radius of $\sim$$27$~kpc \citep{lopez18} with a metallicity gradient between $\sim 0.1-2$~$Z_{\odot}$ from the outermost edge to its centre.
%The Sun is at a distance from the Galactic Centre of about 8~kpcs.
The outer Galaxy (hereafter, OG), usually defined as the portion of the Galactic disk beyond $\sim 11$~kpc,
%\footnote{the Sun is located at $R_{\rm GC}\sim 8$~kpc from the Galactic Center, but the OG is usually considered beyond $R_{\rm GC}\sim 11-12$~kpc, where physical and chemical properties start to diverge significantly from the local ones (Fig.~\ref{fig:comparison}).}, 
is characterized not only by sub-solar metallicities, but also by several other properties that significantly differ from those of the inner Galaxy. 
A scheme showing the environmental differences relevant for star formation between inner and outer Galaxy is shown in Fig.~\ref{fig:comparison}.
In summary, compared with the local Galaxy the OG exhibits: (i) metallicity reduced by factors $\sim 2-10$ depending on the Galactocentric distance, $R_{\rm GC}$ \citep[e.g.,][]{lel11}; (ii) lower dust abundance, and hence gas-to-dust ratios above the $\sim 100$ value inferred locally \citep[e.g.,][]{giannetti17}; (iii) lower gas surface densities \citep[e.g.,][]{hed15,braine23} and ambient pressure \citep[e.g.,][]{wolfire03,schruba19}; (iv) lower star-formation rate \citep[e.g.][]{roman10}.
(v) presumably lower cosmic-ray ionization rates (e.g. Neufeld \& Wolfire~\citeyear{new17}; the local value is $\sim 10^{-17}-10^{-16}$~sec$^{-1}$, Cummings et al.~\citeyear{cummings16}, Sabatini et al.~\citeyear{sabatini23}) and weaker interstellar radiation fields (e.g. Bialy~\citeyear{bialy20}; the local value is $\sim 1$ G$_0$, Habing~\citeyear{habing68}), owing to a reduced amount of sources at the origin of both phoenomena, such as massive stars \citep[e.g.,][]{hunt25} and supernovae.
%, although this point is debated and reliable measurements are still missing.
These properties resemble those found in nearby low-metallicity galaxies like the Magellanic Clouds \citep[e.g.,][]{red92,choudhury18} and, to some extent, galaxies at earlier cosmic epochs. 

Among the nearest laboratories to study in detail the star formation process at sub-solar metallicities, the Large and Small Magellanic Clouds (LMC and SMC, respectively) occupy a prominent place.
The LMC possesses profound differences compared with the OG in the Milky Way.
In addition to having a sub-solar metallicity ($\sim 0.5-0.3~Z_\odot$, Roman-Duval et al.~\citeyear{roman14}),
the intensity of the UV radiation field is higher (10–100 times, but with large variations) when compared with the Galactic values.
The cosmic-ray density is also a factor of 4 lower than that measured in the solar neighborhood \citep[e.g.,][]{sewilo22,shimonishi26}.
Similarly, the SMC has a metallicity of $\sim$$0.25-0.1$~$Z_\odot$, the lowest metallicity of any gas-rich galaxy in close proximity \citep{bolatto07}.
Even if ten times further away than the OG, both the LMC and SMC are sufficiently close ($\sim$$50$~kpc and $\sim$$62$~kpc, respectively) to enable detailed studies of individual star-forming regions.
%\footnote{Star-forming regions are typically decomposed in hierarchical structures from 'clouds' with linear sizes of tens of~pc, to 'clumps' and 'cores' characterised by sizes of $\sim 1$~pc and $\leq 0.1$~pc, respectively \citep[e.g.,][]{motte18}}. 
%Moreover, they do not suffer from distance ambiguities, line-of-sight confusion, and extinction that can hamper in some cases Galactic studies.
In contrast to other galaxies in the local Universe \citep[e.g.][]{rodriguez26}, the OG and Magellanic clouds allow us to resolve individual molecular clouds, prestellar cores, and protostars with unprecedented detail.
{\bf Their proximity and cloud environment properties make the OG and Magellanic clouds a crucial bridge between local, resolved star formation and extragalactic, integrated star formation.}

\vspace{-0.2cm}
\section{The outstanding questions}
\label{question}
\vspace{-0.2cm}
Star formation models generally consider a common set of physical processes —the ``ingredients'' driving the basic physics of star formation— including self-gravity, thermal support, turbulence, magnetic fields, stellar feedback, and others (Fig.~\ref{fig:comparison}).
However, their relative importance may differ between models as these processes depend on the local environment.
In particular, as changes in $Z$ reflect in different coolling and ionisation rates, among others (Fig.~\ref{fig:comparison}), the relative importance of such ``ingredients'' for star formation may vary in low-$Z$ environments with respect to local ones.
Therefore, the fundamental question is: {\bf how do environmental conditions influence the physical and chemical pathways leading to star formation?}
This overarching question raises several specific issues:
(1) Are the properties of prestellar cores universal?
(2) Does cloud fragmentation, and the resulting stellar IMF, depend critically on metallicity?
(3) How does dust evolution depend on metallicity, and what is the impact in the survival of protoplanetary disks?
(4) Is the reduced amount of metals and dust shielding critical for the formation and survival of molecules?

Past and ongoing surveys continue to test whether physical and chemical measurable properties such as the core mass function, the star formation efficiency, and the molecular abundances, vary systematically with Galactic environment (see the ALMA2040 white paper by S\'anchez-Monge et al. 2026).
Classical star-formation theories have posited that the gas reservoirs from which individual stars form are dense, gravitationally bound cloud structures of $1000-10\,000$~au, known as cores \citep[e.g.,][]{mckeeOstriker2007}. 
At the foundation of these theories lies the discovery that the core mass function (CMF) measured in low-mass star-forming regions closely resembles the IMF \citep[e.g.,][]{motte1998,konyves2015}. 
However, one of the transformative results of ALMA, from the ALMA-IMF and ALMAGAL Large Programs, is that the CMF is not universal \citep{motte2018b,coletta2025}. In massive protoclusters, the high-mass end of the CMF ($>$1~$M_\odot$) varies with the evolutionary stage of the cloud, whether quiescent, undergoing a star formation burst, or influenced by feedback from by H\mbox{\sc ~ii} regions \citep{pouteau2023, armante2024, morii2026}. 
While these studies offer physical resolutions that cannot be achieved in other galaxies, they sample only a limited range of environmental conditions. \textit{In particular, the effect of metallicity on cloud fragmentation and the assembly of the CMF remains beyond our current understanding.}

On the 'chemistry' side, several observational efforts have been made to derive molecular abundances in the OG, and investigate possible trends/changes of the chemical composition with $R_{\rm GC}$.
Among these, the 'CHEMical complexity in star-forming regions of the OUTer Galaxy' project \citep[CHEMOUT,][]{fontani22a,colzi22}, which studies the molecular composition of star-forming regions with $R_{\rm GC}$ in the range $\sim$$9-23$~kpc through observations carried out with the IRAM 30m telescope, has indicated that the abundances of most of the molecules studied scale with that of the parent element(s).
This suggests that the molecular formation efficiency is at least as high as in the local Galaxy \citep[e.g.,][]{fontani22b,gigli25}.
This result, in agreement with those of similar observational surveys in the OG \citep[e.g.,][]{bernal21,shimonishi21,wang26} and in low-metallicity nearby galaxies \citep[e.g.,][]{sewilo22,hamedani24,sewilo25}, points to the conclusion that the environmental physical differences may counterbalance each-other to produce a similar chemistry. 
But these results are obtained from selected regions, and \textit{cannot be considered as representative, because systematic surveys in the OG and in the Magellanic clouds, especially at high-angular resolution, are still lacking.}
\vspace{-0.2cm}
\section{Science goals and the need for ALMA2040}
\label{goals}

\vspace{-0.2cm}
\subsection{Starless and prestellar cores at sub-Solar metallicities}
\label{prestellar}
\begin{wrapfigure}[26]{l}{0.65\textwidth}
    \centering
    \includegraphics[width=1\textwidth]{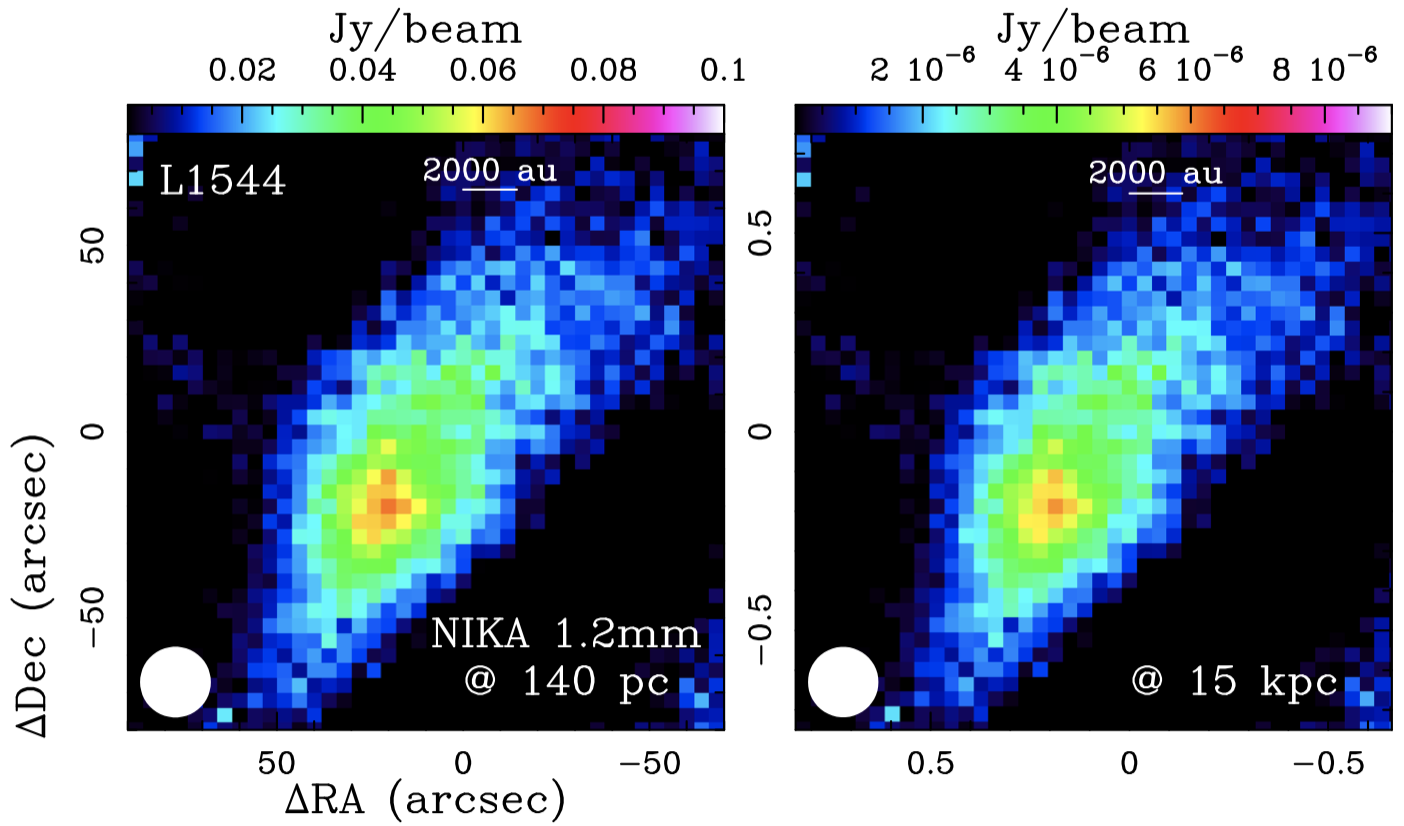}
    \caption{(Left) Map of the continuum emission of L1544 at 1.2 mm obtained with the NIKA receiver of the IRAM 30m telescope by \citet{chacon17}.
A representative linear scale of 2000 au, which corresponds to $\sim 14.3$\arcsec\ at the source distance of 140 pc, is shown.
The ellipse in the bottom-left corner represents the smoothed angular resolution of the image, i.e. $\sim 19$\arcsec.
(Right) synthetic 1.2~mm image of L1544 at a distance of 15~kpc.
The map is obtained with an angular resolution of 0.18\arcsec\ (ellipse in the bottom-left corner), which matches the linear resolution of the original map in the left panel ($\sim 2660$~au).}
    \label{fig:L1544}
\end{wrapfigure}

\vspace{-0.2cm}
Starless cores in clouds of our solar neighborhood are cold overdensities within clouds, with average densities $\geq 10^4$ \cmc, that is about ten times larger than the surrounding medium, and with temperatures of about 10 K.
Prestellar cores are starless cores characterized by higher central densities \citep[$\geq 10^5$ \cmc,][]{aikawa26} which, owing to the high density contrasts between their nuclei and outer edges, are dynamically unstable and tend to gravitationally collapse and form stars.
In prestellar cores, turbulence is subsonic \citep{goodman98} and lines are typically much narrower than 1~$\kms$.
The combination of low temperatures and high densities leads to catastrophic molecular freeze-out in the nucleus of prestellar cores, as (non-thermal) desorption rate of CO can no longer counteract the adsorption rate, which increases with gas density \citep{caselli99}.
CO freeze-out boosts deuterium fractionation, enormously increasing the abundance of deuterated species such ad ${\rm H_2D^+}$, ${\rm N_2D^+}$, ${\rm DCO^+}$, ${\rm NH_2D}$, and others \citep[e.g.,][]{ceccarelli14}.
A prototypical prestellar core is L1544, at a distance of $\sim$140~pc \citep[Fig.~\ref{fig:L1544},][]{chacon17}.
%, a 8 ${\rm M_{\odot}}$ core \citep{tafalla98} lying on the eastern side of the Taurus Molecular 
%At a distance of $\sim$140~pc \citep{chacon17},
%L1544 has a well determined temperature and density profile and a layered chemical structure \citep[e.g.][]{spezzano17}.

A critical difference between starless cores and a prestellar core like L1544 is that the main cooling mechanism in starless cores is CO line emission, while in prestellar cores (at least in the CO-depleted nucleus) is dust thermal emission.
At sub-solar metallicity, both the dust and the CO abundance are expected to decrease, but do they decline proportionally, or do significant differences emerge?
Moreover, is the typical layered chemical structure characterised by high abundance of deuterated species and CO depletion in the nucleus modified in low-metallicity warmer and less dense environments?
So far, physical and chemical properties of prestellar cores have been derived almost totally towards nearby ($d\leq 1$~kpc) star-forming regions \citep[e.g.,][]{andre14,caselli25}.
A primary goal of ALMA2040 will be to accurately \textcolor{blue}{\em determine the physical and chemical structure of a prestellar core like L1544 in the OG, namely at a distance of 10-20~kpc}.
At these distances, both continuum and line emission arising from these cold and line-poor objects are expected to be very faint, as shown in Fig.~\ref{fig:L1544}.
Therefore, the main goal will be achieved thanks to the sensitivity improvements which will enable to study tens of candidates in reasonable amounts of observing times.
A quantitative example is described in Sect.~\ref{technical}).
Moreover, the inner structure of these prestellar cores can be resolved at distances of tens of kpcs only by an interferometer, as their angular size will be $\leq 1$\arcsec, as suggested by Fig.~\ref{fig:L1544}. 
\vspace{-0.2cm}

\subsection{Mass cascade from cloud to cores and the IMF origin at sub-solar metallicity}
\label{fragmentation}
\vspace{-0.2cm}
As shown in \citet{pouteau2023}, the high-mass end of the CMF is rooted in the high-density filamentary cloud structure that surrounds cores. More generally, in gravoturbulent star formation models, cores arise from the fragmentation cascade that develops from cloud to disk scales \citep[e.g.,][]{hennebelle2008, thomasson2024}. This cascade is expected to depend on the balance between turbulence and gravity, as well as non-ideal magnetic field effects, radiative transfer effects, and non-symmetric 3D motions. Observationally, mass cascades, characterized by their fractality index and mass transfer efficiency, continuously vary across the $30\,000-300$~au scale range in protoclusters \citep{thomasson2022, motte2026}. These mass cascades directly trace accretion streams, often funneled by filaments that link scales \citep{pineda2023,hacar23}. At large scales, global motions and turbulence drive the mass cascades, while at $<$$1000-10\,000$~au scales, gravity, rotation, magnetic fields, radiative transfer effect, and stellar feedback shape them
%In the Milky Way, the fragmentation cascade is expected to first change at small scales, in the densest cloud regions corresponding to the core scales ($\sim$1000--10\,000~au), where gravity and optical depth increase, and rotation and magnetic fields intensify, but also
%However, it is also expected to change near newly formed stars, which heat their surrounding gas, drive outflows, and potentially create H\mbox{\sc ~ii} regions 
\citep{yoo2025, motte2026}. 

Radiative transfer effects, in particular the change of equation of state at high densities, stellar heating, and magnetic fields effects are expected to be different in the low-metallicity environments of the outer Galaxy or the Magellanic Clouds. Their combined impact on the mass cascade, the resulting core mass function (CMF), and potentially the IMF itself, remains poorly understood. To date, the only CMF measured in a low-metallicity environment is that of an evolved massive protocluster in the Large Magellanic Cloud, whose cloud is influenced by H\mbox{\sc ~ii} regions \citep{traficante2026}. The high-mass end of its CMF has a slope close to the Salpeter canonical IMF slope, a behavior consistent with CMFs observed in evolved, massive protoclusters of the Milky Way \citep[e.g.][]{armante2024}. 
%but this likely reflects the evolved stage of this massive protocluster rather than a general property of the Large Magellanic Cloud. 
%Indeed, in the inner Milky Way, massive protoclusters under the influence of H\mbox{\sc ~ii} regions consist of low-density clouds that host only low-mass cores and also exhibit a Salpeter-like CMF \citep{armante2024}. 
\textcolor{blue}{\em Statistical investigations of the mass cascades of clouds in low-metallicity environments and their associated CMFs are needed to explore the origin of the IMF in these clouds.}

Studying the fragmentation cascade down to core scales in low-metallicity environments requires observing more distant (at $10-60$~kpc from the Sun), low-dust emitting clouds in the outer Galaxy and Magellanic Clouds, with a wide dynamical range across spatial scales (over $3-4$ decades in scales down to 1000~au). This necessitates significantly higher angular resolution, extremely high continuum sensitivity, and improved image fidelity with high dynamic ranges -far beyond the current capabilities of ALMA (see Sect.~\ref{whyalma}).
\vspace{-0.2cm}
\subsection{Protoplanetary disks at sub-solar metallicity}
\label{disks}
\vspace{-0.2cm}
The formation and evolution of planetary systems are expected to depend strongly on the availability of heavy elements \citep[e.g.,][, see the ALMA2040 white paper by Kuffmeier et al. 2026]{nakatani18,kuffmeier20,andama24} and on the metallicity of their host stars \citep{fev05}. 
Despite recent studies have started to investigate the presence of protoplanetary disks around sub-solar metallicity stars \citep[e.g.,][]{demarchi24,yasui26},
yet most observational constraints on protoplanetary disks come from nearby, approximately solar-metallicity star-forming regions. 
Because a significant fraction of star formation throughout cosmic history occurred in environments with sub-solar metallicities,
establishing how disk properties change with metallicity is essential for determining whether the processes that produce planetary systems 
%in the Solar neighborhood 
were, and still are, universal or environment-dependent.

A central objective of ALMA2040 will be to \textcolor{blue}{\em establish the relation between environmental metallicity and global disk properties, including disk mass, size, dust opacity, gas content, and molecular abundances.}
%By comparing statistically significant samples spanning different metallicities and also different protostellar masses and ages, one could distinguish metallicity-driven effects from those associated with other properties like protostellar mass and disk age. 
%Spatially resolved observations will be particularly powerful in testing whether the reduced solid inventory expected at low metallicity changes the locations and efficiency of dust concentration and planetesimal formation.
In particular, continuum observations will measure the extent and spectral properties of the dust emission, and, for the less distant regions, also the disk structure, while multi-line molecular observations will constrain the bulk temperature, density, kinematics, and chemical composition of the gas. 
Combining these diagnostics will allow us to determine whether the lower metallicity primarily reduces disk masses and dust reservoirs, changes the gas-to-dust ratio, or fundamentally modifies the physical and chemical pathways of disk evolution.
Moreover, local studies indicate a very large variance in disk gas and dust properties \citep[e.g.][]{ansdell16}. 
Comparisons between solar and sub-solar $Z$ studies thus require statistically significant samples to test the robustness of any apparent metallicity trend.
Quantitative examples of observations of protoplanetary disks in the OG that can be carried out with ALMA2040 are described in Sect.~\ref{technical} and in Kuffmeier et al.~(2026).

%Typical integrated flux densities of bright protoplanetary disks are $\geq 20$ mJy in Band 6 and $\geq 100$ in Band 7 \citep[e.g.,][]{villenave20}.
%As an example, Tau 042021 is located at a heliocentric distance of $\sim 130$~pc.
%Its deconvolved angular sizes and integrated flux densities are $\sim 4.1$\arcsec\ and  $F_{\nu}\sim 124$~mJy in Band 7, and $\sim 3.76$\arcsec\ and $F_{\nu}\sim 47$~mJy in Band 6 \citep[e.g.,][]{villenave20}.
%At representative distances of $\sim 10$~kpcs in the OG, the emission size would be $\sim 0.05$\arcsec\, and $F_{\nu}$ would translate into $F_{\nu}\sim 0.02$~mJy and $F_{\nu}\sim 0.007$~mJy in bands 7 and 6, respectively.
%To detect these disks with current ALMA with a signal-to-noise of 4 would require a sensitivity of 0.005 mJy/beam in Band 7, and 0.0015 mJy/beam in Band 6, which can be achieved with an integration time of 30~hrs in Band 7 and several days in Band 6.
%A factor 10 improvement in sensitivity will allow to perform this kind of observations in Band 7 in surveys of objects which, at smaller heliocentric distances, can even allow to perform studies of the emission structure.
\vspace{-0.2cm}
\subsection{Chemical molecular complexity at sub-solar metallicity}
\label{molecules}
\vspace{-0.2cm}
The formation and survival of (complex) organic molecules should
differ in solar and sub-solar metallicity environments due to both the reduced amount of metals and other environmental differences (e.g. UV irradiation from high-mass stars, cosmic-ray ionization rate, e.g.~Bialy \& Sternberg~\citeyear{bes15}).
Nevertheless, as discussed in Sect.~\ref{question}, the molecular abundances investigated so far decrease no more steeply than their corresponding elemental abundances. 

\begin{wrapfigure}[35]{l}{0.58\textwidth}
    \centering
    \includegraphics[width=1\textwidth]{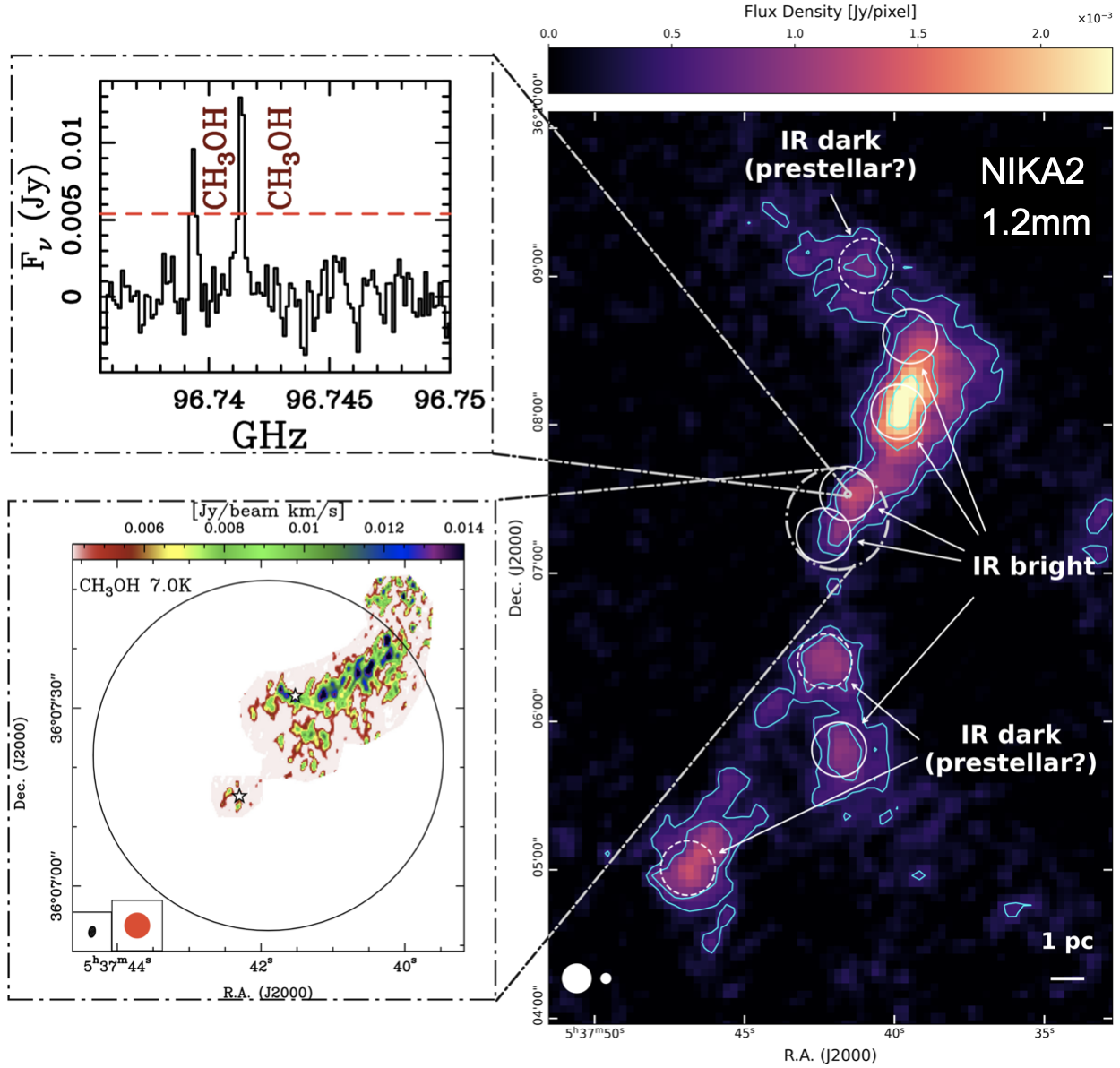}
    \caption{{\it Right}: NIKA2 dust continuum maps of WB89-670 at 1.2 mm \citep{gigli25}. Cyan contours
represent the 1.2~mm emission at 1$\sigma$ (dashed), 3$\sigma$ (solid), 5$\sigma$, 10$\sigma$, and subsequent levels. White circles mark the mid-infrared and millimeter peaks; the solid ones indicate sources with infrared counterparts, while the dashed ones mark peaks without infrared emission. 
In the NIKA2 map, the beam at 150~GHz ($\sim 17$~\arcsec) and a representative 4\arcsec\ beam are illustrated in the bottom-left corner.
{\it Left}: Spectrum and line emission map of methanol obtained with ALMA single-pointing \citep{fontani24}. 
In the map, the 1\arcsec\ beam and a comparative red 4\arcsec\ beam are both shown in the bottom-left corner.}
    \label{fig:wb89-670}
\end{wrapfigure}
Whether this reflects a negligible influence of other environmental factors, or the compensation of competing effects, remains an open question.
To understand this apparently puzzling result, chemical models tailored for the specific conditions of sub-solar metallicity environments, like the OG and the Magellanic clouds, is essential.
However, because star-forming regions (can) have a complex physical structure,
resolving the molecular emission on small angular scales is crucial to perform a meaningful comparison.

As an example, current ALMA has mapped the source WB89-670, located in the OG at $R_{\rm GC}$$\sim 23.4$~kpc \citep{fontani24}, where metallicity is estimated to be $\sim 0.1\, Z_{\odot}$ and the C/O ratio is $\sim 3$ times lower than the solar value according
to Galactic elemental gradients \citep[e.g.,][]{mendez22}.
The source exhibits an internal filamentary structure with cores in various stages (Fig.~\ref{fig:wb89-670}), resembling local counterparts despite its extreme location.
ALMA observations in Band 3 and 4 have provided the first emission maps of multiple transitions of organic species, including methanol (Fig.~\ref{fig:wb89-670}) and carbon-rich species, at such extreme Galactocentric distances \citep{fontani24}, confirming that chemical complexity persists even in the
regions of the Milky Way believed to be the poorest in metals.
However, these maps, obtained with single-pointing observations, provided only a snapshot of a much larger system (Fig.~\ref{fig:wb89-670}).
%Observations of the dust thermal continuum emission at 1.2 and 2 mm \citep{gigli25} have revealed that WB89-670 is part of a large, extended filamentary structure hosting several cores, some of them with infrared counterpart and some without
%(Fig. 1), suggesting the simultaneous presence in the extended structure of both potential starless (prestellar?) and protostellar cores.

The filament hosting WB89-670 resembles filaments 
in the local and inner Galaxy, which are believed to be the pivotal step in the star-formation process \citep[e.g.][]{hacar23,beuther25}; but it is located at the extreme periphery of the Galactic disk. 
The single-pointing observations performed by \citet{fontani24} required 1~hour to 
obtain a sensitivity of $\sim 3.7$~mJy beam$^{-1}$ in a $\sim 0.38$~km s$^{-1}$ channel (Fig.~\ref{fig:wb89-670}), which allowed to
detect methanol with a signal-to-noise ratio of $\sim 5-10$ across the filament.
%Therefore, performing a mosaic of the full region (Fig.~\ref{fig:wb89-670}) to study the filament in its entirety, and in multiple spectral setups, with current ALMA would require much more than $\sim 20$ hours.
%{\bf With current ALMA, mapping the most prominent cores along WB89-670 would require, conservatively, 15 fields in a 3~mm setup (Fig.~\ref{fig:wb89-670}).
%To perform these observations with the same sensitivity and spectral resolution would require $\sim 9-10$~hours.
%Thus, a sample of 10 filaments would need 4 days in just one spectral setup at the same distance of WB89-670.
\textcolor{blue}{\em A sensitivity improvement by a factor 10 will be transformational in the Magellanic Clouds.}
In fact, at their distances ($3-4$ times larger than that of WB89-670), the line intensity should drop by a factor $9-16$.
Considering a factor 10, the sensitivity needed to match that reached by \citet{fontani24} towards WB89-670 is $\sim 0.37$~mJy/beam, feasible with current ALMA in more than one day.
A factor $\geq 10$ improvement in sensitivity, the goal of ALMA2040, will allow to observe tens of filaments like WB89-670.
In fact, in Band 3, a sensitivity of $\sim 0.37$~mJy/beam in a $\sim 0.4$ km s$^{-1}$ bandwidth will be reached in just $\sim 30$ minutes, allowing $10$, or more, sources to be performed in less than 10 hours.

\vspace{-0.2cm}
\subsection{Why ALMA2040?}
\label{whyalma}
\vspace{-0.2cm}
Despite current ALMA capabilities have already enabled groundbreaking studies in the field of star formation
(see Sect.~\ref{question}), low-metallicity environments had lagged behind. 
As discussed above, the OG targets at $R_{\rm GC}$$\sim 15–20$~kpc are intrinsically fainter and smaller in angular size, making orders-of-magnitude improvements in (both line and continuum) sensitivity and angular resolution essential.
The need for higher resolution and sensitivity is even more urgent for the Magellanic clouds, which are $3-4$ times further away than OG sources.
The ALMA2040 capabilities should be able to address the multiple science cases presented above thanks to:
(1) a $10\times$ increase in instantaneous spectral-line sensitivity and continuum over current ALMA, $\sim 8\times$ with respect to ALMA Wideband Sensitivity Upgrade \citep[ALMA-WSU,][]{carpenter23} in lines and $\sim 5\times$ in continuum, to detect weak lines and continuum from prestellar and protostellar cores in such distant low-metallicity environments (see also the ALMA2040 white papers by S\'anchez-Monge et al. 2026, Colzi et al.~2026, and Rivilla et al.~2026);
(2) statistically meaningful surveys of entire molecular cloud complexes with angular size of several arcminutes rather than isolated pointings, as well as surveys of protoplanetary disks with different ages and masses (for details, see also the ALMA2040 white paper by Kuffmeier et al. 2026);
(3) a $\geq 512$~GHz instantaneous bandwidth, allowing simultaneous observations of numerous chemical tracers, eliminating the need for separate observations with different spectral setups;
(4) a $9\times$ smaller PSF area through 3× longer baselines than ALMA-WSU, to resolve prestellar cores, disks, and fragmentation scales (i.e. linear scales comparable to the Jeans length) at distances $\geq 10$~kpc.
%, and to enable direct comparisons between the inner and outer Galaxy under identical linear resolutions.

\vspace{-0.2cm}
\section{ALMA2040: Technical Requirements}
\label{technical}
%Here we summarize the technical requirements to achieve the described science goals.
\vspace{-0.2cm}

\textbf{$\triangleright$ Sensitivity.}
One of the main limitations of current ALMA for the described science cases is the sensitivity to faint continuum sources and lines.
Transitions of a few gaseous species in both prestellar cores and protoplanetary disks can be detected in local star-forming cores.
For example, as discussed in Sect.~\ref{prestellar}, at a representative distance of 15 kpc the prestellar core L1544 would have an angular size of $\sim 0.75$\arcsec, and 
the continuum emission peak at 1.2~mm would be of $\sim 8 \mu$Jy/beam (in a 0.18\arcsec\ beam, Fig.~\ref{fig:L1544}).
In addition, low-J transitions of ${\rm N_2D^+}$, which are 
almost unique tracers of the kinematics in the L1544 nucleus, would appear with peak intensity of 10-30~mK.
These intensity peaks of both continuum and lines would require integration times of days with current ALMA to achieve a signal-to-noise ratio of $\sim 5$.
ALMA2040 would reduce the required integration time to $\sim 1$ hour per source.
In fact, in the continuum this integration time will provide a point-source sensitivity of 1.6 $\mu$Jy in Band 6 (in a 7.5~GHz bandwidth), allowing a core like L1544 to be detected with S/N$\sim 5$ (Fig.~\ref{fig:L1544}), and enabling surveys of tens of sources.
The improved sensitivity will also be essential to detect and study protoplanetary disks in the OG. 
For example, do the masses or sizes of the disks around young stars in the OG differ significantly from those studied in the solar neighbourhood
\citep[see][for a review]{tes24},
%(see Tobin & Sheehan 2024, ARA&A for a review), 
related, for example, to lesser degree of braking due to magnetic fields, turbulence, or irradiation? 
With the increased sensitivity of ALMA2040 it will be possible to obtain $3 \sigma$ sensitivities to 1 $M_{\earth}$ (dust) disks around a 1 $L_{\odot}$ protostar in the OG at 10 kpc in 10 hours of observations at 0.8 mm – comparable to the sensitivity of the survey of disks in Orion by \citet{tobin20}.
Furthermore, the ALMA2040 goal sensitivity will be transformational for the Magellanic Clouds to perform studies of molecular complexity like that presented in Sect.~\ref{molecules}.
%The improved sensitivity will also be essential to detect and study protoplanetary disks in the OG.
%Typical integrated flux densities of bright protoplanetary disks are $\geq 100$~mJy in Band 7 \citep[e.g.,][]{villenave20}.
%As an example, Tau 042021, located at a heliocentric distance of $\sim 130$~pc, 
%in Band 7 has deconvolved angular size and integrated flux density of $\sim 4.1$\arcsec\ and $F_{\nu}\sim 124$~mJy, respectively \citep[e.g.,][]{villenave20}.
%, and $\sim 3.76$\arcsec\ and $F_{\nu}\sim 47$~mJy in Band 6 \citep[e.g.,][]{villenave20}.
%At a representative distance of $\sim 10$~kpc in the OG, the emission size would be $\sim 0.05$\arcsec\, and $F_{\nu}$ would translate into $F_{\nu}\sim 0.02$~mJy in Band 7.
%and $F_{\nu}\sim 0.007$~mJy in bands 7 and 6, respectively.
%To detect this disk with current ALMA with a signal-to-noise of 4 would require a sensitivity of 0.005 mJy, which can be achieved in $\sim 30$~hrs.
%, and 0.0015 mJy/beam in Band 6, 
%which can be achieved with an integration time of 30~hrs.
%A factor 10 improvement in sensitivity will allow to perform these observations in surveys of objects in reasonable integration times.

\textbf{$\triangleright$ Angular resolution and spatial scales.}
Another critical limitation of current ALMA is the angular resolution.
At a representative distance of $15$~kpc in the OG, to resolve linear scales of $10\,000$ au (core envelope scale), 1000 au (core nucleus scale), and 100 au (protoplanetary disk scale) one needs an angular resolution of 0.7, 0.07, and 0.007\arcsec.
An angular resolution of 0.007\arcsec\ is generally unfeasible with current ALMA, and also 0.07\arcsec\ can be reached with extended configurations only.
%that do not operate for long during an observing season, and thus hinder the feasibility of large source surveys.
At the same time, observations should be able to recover the extended emission from molecular clouds and envelopes (see also the ALMA2040 white paper by Rivilla et al. 2026), requiring sensitivity to spatial scales of tens of arcseconds or more, like in the case of WB89-670 (Fig.~\ref{fig:wb89-670}). 
Therefore, the combination of inteferometric capabilities as well as short/zero-spacing observations will remain essential.
Improved image fidelity with high dynamic ranges will be also essential to study the fragmentation cascade (see Sect.~\ref{fragmentation}).

\textbf{$\triangleright$ Bandwidth and multi-transition observations.}
Current ALMA also requires multiple separated spectral setups to cover different rotational transitions of the same molecule.
Covering multiple transitions is essential to derive accurately the excitation temperature of each species and, from this, the molecular column densities.
The WSU will increase the instantaneous bandwidth from the current $\sim16$ GHz total IF bandwidth to $\sim64$ GHz \citep{carpenter23}. 
ALMA2040 will provide broader instantaneous frequency coverage up to 512 GHz, allowing multiple transitions to be observed simultaneously.
For example, low-excitation ${\rm N_2D^+}$ and DCO$^+$ lines, both powerful tracers of starless and prestellar cores, require now separate setups in band 2 and 4 to observe their low-$J$ rotational transitions $J=1-0$ (@$\sim 77.109$~GHz for ${\rm N_2D^+}$, @$\sim 72.039$~GHz for DCO$^+$), and $J=2-1$ (@$\sim 154.217$~GHz for ${\rm N_2D^+}$, @$\sim 144.077$~GHz for DCO$^+$).
In summary, an extension to a bandwidth of 512 GHz will be transformative, allowing simultaneous multi-band observations that will cover the whole series of ground and excited states for all relevant molecules at once, ensuring a common calibration and enabling robust relative comparisons between lines.
%With a $\gtrsim128$ GHz bandwidth, both could be observed simultaneously in both lines in a unique setup.
%The case of ${\rm N_2D^+}$ and DCO$^+$ can be applied to many similar small linear molecules tracers of physics and kinematics of prestellar cores and protoplanetary disks.

\textbf{$\triangleright$ Spectral resolution.}
The science goals target sources that span a wide range of physical conditions, from warm (very) supersonic protostellar cores characterized by linewidths much wider than $1-2$~$\kms$, to cold subsonic prestellar cores in which the lines can be as narrow as $\sim 0.1$~$\kms$.
Flexible spectral-resolution capabilities will therefore allow to address the different science cases across their different physical environments.

%\subsection{Simulations with GHOST}
%Describe the configurations used to obtain the mock observations from GHOST at \url{https://alma-ghost.jb.man.ac.uk} and present the results.
\vspace{-0.2cm}
\section{Synergies with current and future facilities}
\label{synergy}
\vspace{-0.2cm}
%This section is optional. 
The ALMA2040 science goals presented above will greatly benefit from synergies with current and future facilities such as SKAO and the ngVLA, by extending molecular studies into the cm-wave regime. 
SKA-mid will provide access to numerous low-frequency transitions of complex and heavy molecules, like cyanopolyynes, in distant low-metallicity environments such as the OG and the Magellanic Clouds, while its large field of view will enable efficient mapping of extended molecular clouds and filaments with only a few pointings (e.g., \citealt{bianchi2026}).
%This will complement the higher-frequency, higher-resolution view provided by ALMA2040. 
Together, ALMA2040 and SKAO, both operating in the southern hemisphere, will also enable the discovery and characterization of new star-forming sources in the OG that are inaccessible to northern-hemisphere facilities such as the IRAM NOEMA, and VLA. 
In particular, the ngVLA will instead provide high-resolution access (from $\mathrm{mas}$ to sub-$\mathrm{mas}$ scales) to low-frequency molecular emission, probing deeply embedded protostellar regions and the innermost, planet-forming regions of disks, where dust opacity can limit (sub-)mm observations (e.g., \citealt{oberg2018}). 
%Together, these facilities will provide complementary frequency coverage, field of view, and spatial resolution, from molecular clouds to planet-forming regions.

\vspace{-0.2cm}
\section*{Acknowledgements}
\vspace{-0.2cm}
This work was developed as part of the broader ALMA2040 community effort. We acknowledge the Lorentz Center and NOVA for supporting the community workshops and coordination activities that helped initiate and shape the ALMA2040 White Paper process. We express our gratitude to A. Baudry, L Boogaard, and \'A. Hacar for their valuable comments and suggestions.\\

\vspace{-0.2cm}
\bibliographystyle{apj_mod}
{\bibliography{bibliography}}
\end{document}